\documentclass[
]{ceurart}

\usepackage{listings}
\usepackage{todonotes}
\usepackage{booktabs}
\usepackage{siunitx}
\usepackage{cprotect}
\begin{document}

\copyrightyear{2023}
\copyrightclause{Copyright for this paper by its authors.
  Use permitted under Creative Commons License Attribution 4.0
  International (CC BY 4.0).}

\conference{CLEF 2023: Conference and Labs of the Evaluation Forum, 
    September 18--21, 2023, Thessaloniki, Greece}

\title{Jean-Luc Picard at Touché 2023: Comparing Image Generation, Stance Detection and Feature Matching for Image Retrieval for Arguments}
\title[mode=sub]{Notebook for the Touch{\'e} Lab on Argument and Causal Retrieval at CLEF 2023}

\author[1]{Max Moebius}[%
email=max.moebius@uni-jena.de,
url=https://github.com/ArcticF0x99,
% degree= B.Sc.,
]
\fnmark[1]
\cormark[1]

\author[1]{Maximilian Enderling}[%
email=maximilian.enderling@uni-jena.de,
url=https://github.com/BMI24,
% degree= B.Sc., 
]
\fnmark[1]

\author[1]{Sarah T. Bachinger}[%
orcid=0009-0005-5422-2164,
email=sarah.bachinger@uni-jena.de,
url=https://github.com/stbachinger,
% degree= B.Sc.,
]
\fnmark[1]

\address[1]{Friedrich-Schiller-University Jena,
  07737 Jena, Germany}

\cortext[1]{Corresponding author.}
\fntext[1]{These authors contributed equally.}

\begin{abstract}
  Participating in the shared task "Image Retrieval for arguments", we used different pipelines for image retrieval containing Image Generation, Stance Detection, Preselection and Feature Matching. We submitted four different runs with different pipeline layout and compare them to given baseline. Our pipelines perform similarly to the baseline.
\end{abstract}



\maketitle

\section{Introduction}
As the saying goes, "a picture is worth a thousand words". A convincing argument in writing should be accompanied by an equally convincing image. There are no perfect out-of-the-box solutions so far, which is why we participated in the shared task "Image Retrieval for arguments".

\section{Related work}
In the following section, related work covering image generation with Stable Diffusion and Feature Matching is reviewed.

\subsection{Stable Diffusion}

''Stable Diffusion is a latent text-to-image diffusion model capable of generating photo-realistic images given any text input.''\cite{stable}
\\
\\
Stable Diffusion is a neural network that generates an image corresponding to a given text input (so-called \textbf{prompt}). If additionally a style, like ''comic'' is included in the prompt, the image is generated in this specific style.
\\
\\
For our approach, the version \texttt{stable-diffusion-v1-4} was used, which was created by resuming training from stable-diffusion-v1-2 with 225,000 steps at a resolution of 512x512 pixels\cite{stable}.

\subsection{Feature Matching}

''Feature matching refers to the act of recognizing features of the same object across images with slightly different viewpoints.''\cite{feature} In this context, features are defined by keypoints, which causes their name \texttt{feature keypoints}. These feature keypoints mark an area that is particularly interesting or defining in an image.
\\
\\
\texttt{SIFT} is a feature descriptor used to detect, describe and match local features of images. For that, the descriptor uses a database of images to compare with. Every feature of the new image is compared to the database with Euclidean distance of the feature vectors to recognize objects.\cite{siftWiki} Broadly speaking, SIFT extracts feature keypoints and feature descriptor from an image. The descriptors contain the visual description of images and are usually used to determine the similarity between images.
\\
\\
\texttt{FLANN} stands for \texttt{Fast Library for Approximate Nearest Neighbors} and is used for fast nearest neighbor search in large datasets or images. In general, a matcher takes the descriptors of two images, builds pairs of features with one feature from each image, and calculates for every pair a distance. The smaller the distance, the more similar the features are to each other. With clustering and search in multidimensional spaces, the matching by FLANN is more efficient for larger datasets compared to the usually used \texttt{BFMatcher}.\cite{featureMatching}
\\
\\
Using a threshold, the feature matches are filtered. The matches with a distance under the threshold are determined as good and the better the matching between two images, the more similar the images are to each other.
\\
\\
Additionally, \texttt{homography} is used to determine the transformation between points in an image and projects them on to an image plane with a normalized camera. That means objects are viewed usually from different angles in two images, but they show the same objects. Therefore, the images have a different perspective. With homography, the objects in the image are made comparable by bringing the images into the same perspective. (Compare \cite{homography})

\section{Approach}\label{approach}
\begin{figure*}
    \centering
    \includegraphics[width=\textwidth]{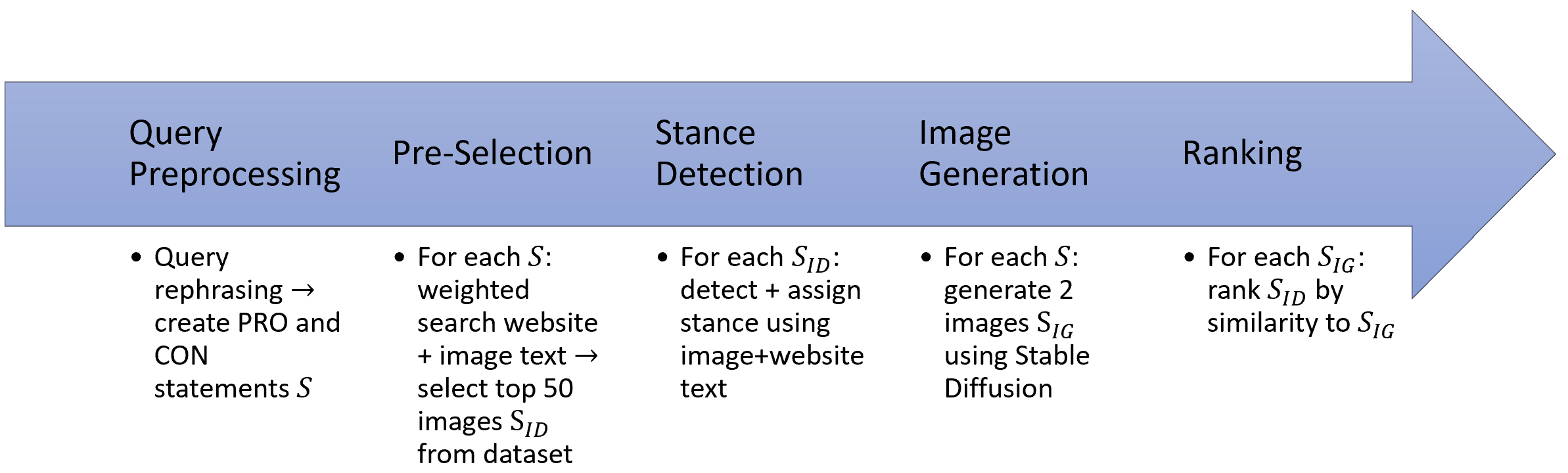}
    \caption{The whole pipeline}
    \label{fig:pipeline}
\end{figure*}

Our implementation is available on GitHub\footnote{\url{https://github.com/ArcticF0x99/ir_image_retrieval}}. The complete pipeline containing all steps is shown in Figure \ref{fig:pipeline}. In its current state, it's a full-rank pipeline.
\subsection{Query Preprocessing}\label{preprocessing}
To isolate possible important terms for the following queries, the spaCy\cite{spacy} library was used to parse the topic questions and generate a parse tree. \\
Then, all tokens that were identified as punctuation and the root word (a verb) were excluded. From the remaining, only words that had a lower Zipf frequency\cite{zipf} than 5.6 were kept. This threshold was chosen because it was the average Zipf word frequency in the given corpus. After this, it was assumed that only relevant words were kept. To support the query, the root word was placed in front of the remaining words, which kept their order. \\\
For example, the phrase "Do we need sex education in schools?" was decomposed into the root word "need" and the remaining words "sex", "education", and "schools". The resulting intermediate query would be "need sex education schools". \\
For PRO queries, the intermediate query was used as it is. For the CON queries, the negation "not" was put in front of the intermediate query.

\subsection{Image Preselection}
% \todo{mention possibility to add rerank}
For the initial image preselection, an index was constructed with PyTerrier\cite{pyterrier} containing the ID of a document and its text content. Here, at most 4096 characters were used.
Then, BM25\cite{bm25} was used to retrieve the best 50 images for a given query, generated as described in \ref{preprocessing}.

\subsection{Stance Detection}
For the stance detection, we used a Hugging Face pipeline\footnote{\url{https://huggingface.co/facebook/bart-large-mnli}} that implements the work from \citet{zeroshotClassification} for zero-shot classification with the BART model after being trained on the MultiNLI data set. The pipeline was freely available and can be loaded with the ''zero-shot-classification" pipeline from hugging face\footnote{classifier = pipeline("zero-shot-classification", model="facebook/\allowbreak bart-large-mnli")}.\\
The model receives text and labels and calculates accordingly the probability of the label being a good descriptor for the given text. Hence, either "contra", "pro" or "neutral" was added in front of the given query.\\
The highest probability was assumed to show the stance of the image. They were sorted according to the score and the image IDs returned.

\subsection{Image Generation}

The image generation is used to generate image for given queries/phrases. For that, we used \texttt{Stable Diffusion}.
\\
\\
A generated image for a query should display the information of the query visually. This generated image is then used to compare to other images, like with \texttt{Feature Matching}, to know how similar the images are to each other. This, in turn means, that for every image, a similarity to the query is calculated.

\subsection{Image Ranking (Feature Matching)}

\texttt{Feature Matching} is used to rank a set of images for a given query. The images in the set can have different styles, so feature matching with generated images of different style can give a better overall matching for a query. Therefore, the query is used to generate a photorealistic image and an image in comic design. With feature matching for every image of the set and the generated images, matches are calculated and the number of good matches is returned. So for every image, the count of good matches for every generated image is returned and then summed up. The more good matches an image has, the more it fits to the query. With the number of matches, a ranking of the images is created and then returned.

\section{Results}\label{results}
\subsection{Submission}
We submitted 5 runs with different combinations of the approaches described in Section \ref{approach}, namely:
\begin{enumerate}
    \item \textbf{Pipeline -1}: Baseline provided by the Touché Team
    \item \textbf{Pipeline 0}: Query Preprocessing and Image Preselection only
    \item \textbf{Pipeline 1}: Query Preprocessing and Image Preselection with Stance Detection on text
    \item \textbf{Pipeline 2}: Query Preprocessing and Image Preselection with Stance Detection on image text
    \item \textbf{Pipeline 3}: Query Preprocessing and Image Preselection with Stance Detection on text and image text
\end{enumerate}
In every approach, Image Generation and Image Ranking was used to determine the final ranked results.

\subsection{Relevance evaluation}
From the 5 pipelines, we collected the top 10 images returned by the pro and con queries for each of the 50 topics, gathering a total of 5000 images. After duplicate removal, 1938 images remained and were independently judged by three annotators with the labels \textit{off-topic}, \textit{pro}, \textit{con} and \textit{neutral}. The Fleiss-Kappa for the tree annotators was 0.37.
For the evaluation of the different approaches, the image judgments were curated as follows: 
\begin{itemize}
    \item if two annotators agree on a label, this labels is chosen
    \item if there is no majority agreement and the image is labeled by more than one as \textit{on topic}, its label will be \textit{neutral}
\end{itemize}

\subsection{Pipeline Evaluation}
The returned images for the pipelines were evaluated using the curated judgements and precision was calculated as precision@10, precision@1, and mean average precision (see Table \ref{tab:precision}). Furthermore, we used a student's t-test to find out whether the difference in values for average precision of the individual runs is significant compared to the -1 (baseline) run. The p-values are also available in Table \ref{tab:precision}.
\begin{table}[]
    \centering
    \begin{tabular}{SSSSS}\toprule
    {Pipeline ID} & {Precision@10} & {Precision@1} & {MAP} & {p-value} \\\midrule
    -1 &  0.147 & 0.14 & 0.147 & \\
    0 &  \textbf{0.15} & \textbf{0.18} & 0.15 & 0.916 \\
    1 & 0.13 & 0.08 & 0.132 & 0.598\\
    2 & 0.115 & 0.15 & \textbf{0.185} & 0.185 \\
    3 & 0.148 & 0.07 & 0.155 &0.775\\\bottomrule
    \end{tabular}
    \caption{Precision@10, precision@1, mean average precision (MAP), and p-value for student's test for the pipelines with the curated data corpus}
    \label{tab:precision}
\end{table}

\section{Discussion}\label{discussion}
The results above indicate that even though the precision values vary across the different pipelines, none are statistically significant different from the baseline.\\
We observed relatively low values for precision on average. This may be partially due to missing relevant pictures in the corpus. Judging by the low Fleiss-Kappa, the task of evaluating the stance and relevance of an image for a certain topic was ambiguous, which was confirmed by the annotators. For future applications, an annotation guideline should be given to the annotators to avoid confusion. \\
We see that the inclusion of stance detection on the website text seems to be not beneficial in our case. Future work could determine if other stance detection models also suffer from this. \\
Furthermore, selecting a higher amount of pictures than the current number of 50 during pre-selection could lead to different results. This is supported by the fact that out of the 5000 results for the 5 pipelines, actually only 1938 were unique. Since all of our results are chosen from the same $50$ (\# topics) $\cdot2$ (\# of stances) $\cdot50$ (\# of pre-selected images per query)$=5000$ images, this may be part of the poor results.

\section{Conclusion}
As seen in Section \ref{results}, our pipelines preform not significantly better or worse than the baseline. Using more information, a pipeline should get better than the previous one. So the best pipeline should have been pipeline 3. In our case, the pipeline 0 was the best in precision@10 and precision@1 and pipeline 2 in MAP. \\

Different things can be changed to hopefully get better results. Testing other variations of stance detection and image generation with image ranking. Other, untested approaches and a change in the approach combination used may also help. As mentioned in the section \ref{discussion}, maybe the corpus is at fault and more pictures with a clearer stance may lead to better results. Increasing the number of pre-selected images is very important. 
\bibliography{sample-ceur}

\begin{acknowledgments}
We thank Prof. Hagen and his chair for their continued helpful suggestions and support.
\end{acknowledgments}

\end{document}